\documentclass{article}
\usepackage{authblk}
\usepackage{amsmath}

\title{The gauge-invariant formulation of the local expansion rate driven by the local average density in an inhomogeneous universe}

\author[1]{Masanori Tomonaga \thanks{email:i2185088@cc.kyoto-su.ac.jp}}
\author[2]{Masumi Kasai \thanks{email:kasai@hirosaki-u.ac.jp}}
\author[1]{Toshifumi Futamase \thanks{email:tof@cc.kyoto-su.ac.jp}}

\affil[1]{Faculty of Science, Kyoto Sangyo University, Motoyama, Kamigamo, Kita-ku, Kyoto, 603-8555 Japan}
\affil[2]{Graduate school of Science and Technology, Hirosaki University,Hirosaki 036-8561, Japan}

\date{}

\begin{document}
\maketitle

\begin{abstract}
    The Hubble tension casts a blight on the standard cosmology.
     As a possible solution to the problem, the local variation of the
  expansion rate has been proposed where the spatial averaging over a
  finite domain was introduced in order to restore the local Friedmannian
  behavior in an inhomogeneous cosmology.
     So far, however, the approaches are limited to the particular choices
  of the gauges, and it has been unclear whether the results are
  gauge-invariant.
     In this paper, we present the gauge-invariant formulation of the
  local expansion rate  which is driven by the spatial average of the
  gauge-invariant inhomogeneous density.
     We show that the local cosmological parameters in the finite domain may
     change from the global parameters, and the relations between them are
  expressed by the gauge-invariant averaged density.
  \end{abstract}

\section{Introduction}
The Hubble constant $H_0$ is one of the most important cosmological parameter since it characterizes the global properties of our universe.
The standard cosmology is based on the assumption of the homogeneity and isotropy. 
Thus, the Hubble parameter $H_0$ is regarded as a constant over at least the horizon scale which is also the prediction of the inflationary scenario.
However recent observations suggest a non negligible difference between local and global (or recent and old) Hubble parameter \cite{2020,riess20162}. \\
There has been a large number of studies which try to resolve the discrepancy \cite{ichiki2016relationship,bolejko2018emerging,tomita2017cosmological,tomita2018super,benetti2019looking,akarsu2019constraints,cea2022ellipsoidal,blinov2020warm,bento2002generalized,2021realm,2021darksector,berghaus2020thermal,jedamzik2020relieving,macpherson2018trouble,cai2021chameleon,martin2021hubble}.
We regard that  the difference of the local and global Hubble parameter is real and be
explained  by the inhomogeneous distribution of the matter. 
In fact, the observation of the K-band luminosity density seems to suggest that region with several hundred Mpc
around us has low density with density contrast $\delta_K\sim -0.5$ compared with
the globally averaged density \cite{keenan2013evidence}. Furthermore, there is some indication that the voids are actually low density by weak lensing observation.
Thus, it will be meaningful to pursue the indication of the cosmological inhomogeneity.

The homogeneous and isotropic universe (here we call Friedmann Universe)  appears as
the  result of some kind of averaging procedure since the universe is actually very inhomogeneous. 
There are various ways to averaging inhomogeneous universe (such as the light-corn averaging that is directly related with observational quantities). In this paper, we only consider the scalar perturbations in the linear order and the spatial averaging 
\cite{buchert2002regional,wiltshire2007cosmic,buchert1995averaging,ben2012backreaction,gasperini2011light,fanizza2020generalized,yoo2017gauge,korzynski2010covariant,zalaletdinov2004spacetime,clarkson2012observational,coley2006averaging,heinesen2019covariance,buchert2001average,buchert2000average,boersma1998averaging,PhysRevD.103.083507}.
Our purpose is not studying the averaging itself in general inhomogeneous spacetimes, but rather  the gauge dependence of the relationship between locally averaged and globally averaged spacetime in the linearly perturbed universe using the spatial averaging.
By adopting the spatial averaging defined below, we were able to derive a locally averaged Friedmann universe and have obtained the following  relation between the locally average Hubble parameter and the globally averaged Hubble parameter within the framework of the general relativistic perturbation theory\cite{kasai2019possible,tomofuta}
\begin{equation}
    H_{D0} = H_0\left(1 - \frac{1}{3}f(t_0)\langle\varDelta\rangle_{Dt_0}\right),
\end{equation}
where $H_{D0}$ is the averaged Hubble parameter at the present time $t_0$ over a finite domain $D$, and $H_0$ is the global, or the horizon scale Hubble parameter, and $\langle\varDelta\rangle_{Dt_0}$ is the present density contrast average over the domain $D$. $f(t)=d\log \varDelta/d\log a$ is the growth function of the density contrast.
However, the treatment is carried out in the comoving synchronous and Newtonian gauge, and there is some question if the averaging and the result are gauge invariant or not.

In order to answer the question, we study the spatial averaging in the framework of the gauge invariant cosmological perturbation theory, and find that local Hubble parameter (and local cosmological parameter) can be describe used by gauge-invariant physical quantities averaged in the local region $D$.

%%% Kasai

\section{Gauge-invariant linear perturbation theory}

In this section, we briefly summarize the gauge-invariant perturbation theory \cite{bardeen, kodasasa}. 
We assume the flat background with dust fluid. Then the background metric is
\begin{equation}
  ds^2 = - dt^2 + a^{2}(t) \delta_{ij} dx^i dx^j \, 
\end{equation}
and the energy momentum tensor is 
\begin{equation}
  T^{\mu} = \rho_b(t) \,u^{\mu} u^{\nu}, \quad u^{\mu} = (1, 0, 0, 0) \,. 
\end{equation}

From the Einstein equations, we obtain the following Friedmann equation 
\begin{equation}\label{eq:Friedmann}
  \left(\frac{\dot{a}}{a}\right)^2 = \frac{8\pi G}{3} \rho_b  + \frac{\Lambda}{3}
\end{equation}
and from $T^{\mu\nu}_{\ \ \ ;\nu} = 0$, we obtain the energy conservation equation 
\begin{equation}\label{eq:dotrhob}
  \dot{\rho}_b + 3 \frac{\dot{a}}{a} \rho_b = 0\, .
\end{equation}

Next, we write the metric and the energy-momentum tensor in the perturbed universe as follows:
\begin{align}
    ds^2 &= g_{\mu\nu} dx^{\mu} dx^{\nu} \\
    g_{00} &= - (1 + 2 A) \\
    g_{0i} &= - B_{, i} \\
    g_{ij} &= a^2(t) \left(\delta_{ij} + 2 E_{,ij} + 2 F \delta_{ij} \right) \\
    T^{\mu\nu} &= \rho \, u^{\mu} u^{\nu} \\
    \rho &= \rho_b\,(1 + \delta) \\
    u^{\mu} &= (u^0, u^i) = \left( 1- A, a^{-2} \delta^{ij} v_{,j} \right) , 
\end{align}
where we consider only scalar perturbations and the scalar perturbation variables $A, B, E, F, \delta$ and $v$ are arbitrary functions of $t$ and $x^i$, and assumed to be small quantities.  

Now consider the scalar type infinitesimal gauge transformation 
\begin{align}\label{eq:gaugetr1}
  \bar{t} &=t+\alpha \, ,\\
  \bar{x}^i &= x^i + \delta^{ij} \beta_{,j} \ ,\label{eq:gaugetr2}
\end{align}
where $\alpha$ and $\beta$ are arbitrary functions of $t$ and $x^i$, which are regarded as small as the perturbation variables. 
The gauge dependence of the perturbed quantities are 
\begin{align}
    \bar{A} &= A - \dot{\alpha} \, ,\\ 
    \bar{B} &= B - \alpha + a^2\dot{\beta} \, ,\\
    \bar{E} &= E - \beta \, ,\\ 
    \bar{F} &= F - \frac{\dot{a}}{a}\alpha \, ,\\
    \bar{\delta} &= \delta + 3\frac{\dot{a}}{a}\alpha \, ,\\
    \bar{v}  &= v + a^2\dot{\beta}\, .
\end{align}
Then, the following gauge invariant quantities are defined in the usual manner.
\begin{align}
    \varPhi &\equiv A - \left(B + a^2\dot{E} \right)^{\cdot} \, ,\\
    \varPsi &\equiv F - \frac{\dot{a}}{a} \left(B +a^2\dot{E} \right) \, ,\\
%   \color{red}{\varDelta} &\equiv \color{red}{\delta + 3 \frac{\dot{a}}{a}\left(B +\dot{E}\right)} \, ,\\
{\varDelta} &\equiv \delta - 3 \frac{\dot{a}}{a}\left(v -B\right) \, ,\\
    V &\equiv v + a^2\dot{E}\, .
\end{align}
% All of the above gauge-invariant variables are reduced to the perturbed quantities themselves in the Newtonian gauge $B = E = 0$. 
% Therefore, $\varPhi, \varDelta$, and $V$ correspond to the Newtonian potential, the density contrast, and the perturbed velocity, respectively in the Newtonian gauge. 
% Moreover, the 3-curvature of the $t=\mbox{const.}$ hypresurface is 
% \begin{equation}
%   {}^{(3)}\! R = - \frac{4}{a^2}\nabla^2 F \, ,
% \end{equation}
% where $\nabla^2 \equiv \delta^{ij} \partial_i \partial_j$ is the Laplacian in the background flat space, $\varPsi$ corresponds to the  the 3-curvature of the $t=\mbox{const.}$ hypresurface in the Newtonian gauge. 

% We can define another set of the gauge-invariant variables: 
% \begin{align}
%   \color{red}{\varPhi_m} &\equiv A + \left(\dot{v} - \dot{B}\right) 
%              = \varPhi + \dot{V}\, ,\\
%   \color{red}{\varPsi_m} &\equiv F + \frac{\dot{a}}{a} \left(v - B\right) 
%              = \varPsi + \frac{\dot{a}}{a} V\, ,\\
%  \color{red}{\varDelta_m} 
%    &\equiv \color{red}{\delta - 3 \frac{\dot{a}}{a}\left(v -B\right)} 
%     = \varDelta - 3 \frac{\dot{a}}{a} V \, .\\
% \end{align} 
% From the above definitions, it is apparent that the gauge-inraviant variables 
% $\varPhi_m, \varPsi_m$, and $\varDelta_m$ correspond to the Newtonian potential, the 3-curvature, and  the density contrast in the comoving gauge $v = B$. 

Using these quantities, we can obtain the first-order equations in terms of the gauge invariant quantities of linearized Einstein equation as follows:
\begin{align}
     - \frac{1}{a^2} \nabla^2\varPsi &= 4\pi G \rho_b {\varDelta} \label{00inv}\\
    \frac{\dot{a}}{a}\varPhi - \dot{\varPsi} &= -4\pi G \rho_b V \label{0iinv}\\
    \varPsi + \varPhi &= 0 . \label{ijinv}
\end{align}
 
Using \eqref{ijinv}, the equations \eqref{00inv} and \eqref{0iinv} are re-written as 
\begin{align}
  \frac{1}{a^2} \nabla^2\varPhi &= 4\pi G \rho_b {\varDelta} , \label{00inv2}\\
  \dot{\varPhi} + \frac{\dot{a}}{a}\varPhi  &= -4\pi G \rho_b V . \label{0iinv2}
\end{align}

From $T^{\mu\nu}_{~~;\nu}=0$, we obtain 
\begin{align}
    \dot{\varDelta} + \frac{1}{a^2}\nabla^2 V &= 0, \label{0ec}\\
    \dot{V} + \varPhi &= 0 . \label{iec}
\end{align}
Differentiating \eqref{0ec} with respect to $t$ and using \eqref{00inv2} and \eqref{iec}, we obtain 
\begin{equation}\label{eqddotDelta}
    \ddot{{\varDelta}} + 2\frac{\dot{a}}{a}\dot{{\varDelta}} - 4\pi G \rho_b {\varDelta} = 0.
\end{equation}
The solution of the second-order differential equation \eqref{eqddotDelta} generally has two independent modes as follows:
\begin{equation}
  {{\varDelta}}(t, x^i) = \mathcal{D}_{+}(t) Q_{+}(x^i) + \mathcal{D}_{-}(t) Q_{-}(x^i) \, ,
\end{equation}
where 
\begin{align}
  \mathcal{D}_{+}(t) & = H\int^t \frac{dt'}{\left(a H\right)^2} \, ,\\
  \mathcal{D}_{-}(t) & = H  = \frac{\dot{a}}{a}\, ,
\end{align}
$Q_+(x)$ and $Q_-(x)$ represent the spatially dependent part of the growing and decaying mode of the density contrast, respectively.

In summary, from \eqref{eq:Friedmann} and \eqref{00inv2} multiplied by $2/3$,  we obtain the following equation 
\begin{equation}\label{eq:perF}
  \left(\frac{\dot{a}}{a}\right)^2 + \frac{2}{3} \frac{1}{a^2} \nabla^2 \varPhi 
   = \frac{8\pi G}{3} \left(\rho_b + \rho_b{{\varDelta}}\right) + \frac{\Lambda}{3}
\end{equation}
as the perturbed version of the Friedmann equation, 
and from \eqref{eq:dotrhob} including $\rho_b\varDelta$, 
\begin{equation}\label{eq:perdotrho}
  \frac{\partial}{\partial t} \left(\rho_b + \rho_b{{\varDelta}}\right)
  + 3 \frac{\dot{a}}{a}  \left(\rho_b + \rho_b{{\varDelta}}\right) 
  - \rho_b \dot{\varDelta} = 0
\end{equation}
as the perturbed version of the energy conservation equation.

\section{Spatial averaging over a local domain in the perturbed universe}

In the previous section, we have employed the standard assumption that the inhomogeneous matter density $\rho$ can be decomposed into the homogeneous background part $\rho_b(t)$ and the small perturbed part $\delta$. 
In the actual inhomogeneous universe, however, we need to extract the homogeneous part through the averaging procedure. 

We define the spatial volume $V_D$ of a finite small domain $D$ in the $t = \mbox{const.}$ hypersurface $\Sigma_t$ as 
\begin{equation}\label{eq:VD}
  V_D \equiv \int_D \sqrt{\det(g_{ij})}\, d^3x\,. 
\end{equation}
$D$ is sufficiently smaller than the horizon scale but more than the scale at which the picture of the Hubble expansion is valid, e.g. more than several 10 Mpc.
Using the metric described in the previous section,
$\Sigma_t$ is specified by the normal vector
\begin{equation}
n^{\mu} = \left(1 - A, \frac{1}{a^2} \delta^{ij} B_{,j} \right)\,.
\end{equation}
Contrary to those in \cite{kasai2019possible} and \cite{tomofuta}, no gauge-fixing is made in \eqref{eq:VD} in order to specify the $t = \mbox{const.}$ hypersurface $\Sigma_t$.  
Fixing the gauge $A=B=v=0$ reproduces the results in \cite{kasai2019possible}, and
another gauge $B = E = 0$ leads to those in \cite{tomofuta}. The spatial average of a scalar quantity $Q(t,x^i)$ over the domain $D$ is in general 
\begin{equation}
  \langle Q \rangle \equiv \frac{1}{V_D} \int_D Q \sqrt{\det(g_{ij})}\, d^3x\,.  \label{Qaverage}
\end{equation}
Therefore, the average density in this domain is 
\begin{equation}
  \langle \rho \rangle \equiv \frac{1}{V_D} \int_D \rho \sqrt{\det(g_{ij})}\, d^3x\,.  \label{rhoaverage}
\end{equation}
Since we can observe only a finite nearby region of the entire space, it is likely that the average density $\langle \rho \rangle$ in the nearby region does not always coincides with the background density $\rho_b$.  

Spatially averaging \eqref{eq:perdotrho} over a local domain $D$, we obtain
\begin{equation}\label{eq:perdotrho2}
  \left\langle\frac{\partial \varrho}{\partial t} \right\rangle
  + 3 \frac{\dot{a}}{a}  \langle \varrho \rangle 
  - \rho_b \langle \dot{\varDelta} \rangle = 0\, ,
\end{equation}
where we have defined the gauge-invariant inhomogeneous density 
\begin{equation}
  \varrho \equiv \rho_b + \rho_b \varDelta
\end{equation}
in order to distinguish it from $\rho = \rho_b + \rho_b \delta$. 

Note that the time derivative does not commute with the spatial averaging in general.
In fact, for a physical quantity $Q$ we have
\begin{equation}
  \left\langle\frac{\partial Q}{\partial t} \right\rangle
  - \frac{d}{dt} \left\langle Q \right\rangle = 
  \left\langle \frac{1}{2} g^{ij} \dot{g}_{ij}\right\rangle \left\langle Q \right\rangle 
  - \left\langle \frac{1}{2} g^{ij} \dot{g}_{ij} Q\right\rangle \,. 
\end{equation}
However, if we consider the case $Q \to \rho$ up to the linear order, we obtain
\begin{equation}
  \begin{split}
    \left\langle \frac{\partial\varrho}{\partial t} \right\rangle
    - \frac{d}{dt} \left\langle\varrho\right\rangle
    &=
    \left\langle \frac{1}{2} g^{ij} \dot{g}_{ij}\right\rangle
    \left\langle \rho_b + \rho_b \varDelta \right\rangle
    - \left\langle \frac{1}{2} g^{ij} \dot{g}_{ij} (\rho_b + \rho_b
    \varDelta)\right\rangle \\
    &=
    \left\langle 3\frac{\dot{a}}{a}\right\rangle \left\langle \rho_b
    \varDelta \right\rangle
    - \left\langle 3\frac{\dot{a}}{a} \rho_b \varDelta\right\rangle \\
    &= 0.
  \end{split}
  \label{rhocommute}
\end{equation}
Therefore, using the relation \eqref{rhocommute}, it is straightforward to show from \eqref{eq:perdotrho2} the following equation holds up to the linear order of the perturbations:
\begin{equation}
  \label{eq:perdotrho4}
  \frac{d}{dt}\left\langle\varrho \right\rangle
  + 3 \frac{\dot{a}_D}{a_D} \left\langle \varrho \right\rangle 
  = 0 \,,
\end{equation}
where 
\begin{equation}\label{eq:aD}
  \frac{\dot{a}_D}{a_D}\equiv 
  \frac{\dot{a}}{a} - \frac{1}{3} \langle\dot{\varDelta} \rangle
\end{equation}
can be regarded as the local expansion rate driven by the local average density $\langle \varrho \rangle$. 

% \begin{equation}\label{eq:perdotrho3}
%   \frac{d}{dt}\left\langle\varrho \right\rangle
%   + 3 \left( \frac{\dot{a}}{a} - \frac{1}{3} \langle\dot{\varDelta} \rangle \right)\left\langle \varrho \right\rangle 
%    = 0 \,.
% \end{equation}
% Since we consider only the growing mode solution, 
% \begin{align}
%   \frac{d}{dt}\left\langle\varDelta\right\rangle &=
%   \frac{\dot{a}}{a} \frac{d \ln \mathcal{D}_{+}}{d \ln a}\left\langle\varDelta\right\rangle 
%   = \frac{\dot{a}}{a} \left\langle\varDelta\right\rangle \,, \quad\because \mathcal{D}_{+} \propto a\,.
% \end{align}

In order to express \eqref{eq:perF} in terms of $a_D$, we rewrite as 
\begin{equation}\label{eq:perF2}
  \left(\frac{\dot{a}}{a}\right)^2 - \frac{2}{3} \frac{\dot{a}}{a} \dot{\varDelta} 
  + \frac{2}{3} \frac{1}{a^2} \left(\nabla^2 \varPhi + a^2 \frac{\dot{a}}{a} \dot{\varDelta}\right)
   = \frac{8\pi G}{3} \left(\rho_b + \rho_b{{\varDelta}}\right) + \frac{\Lambda}{3}.
\end{equation}
Spatially averaging \eqref{eq:perF2} and substituting \eqref{eq:aD}, we obtain
\begin{equation}
  \left(\frac{\dot{a}_D}{a_D}\right)^2 +  \frac{K_{\rm eff}}{a_D^2} 
  = \frac{8\pi G}{3}\langle\varrho \rangle + \frac{\Lambda}{3}\, , 
\end{equation}
where
\begin{align}
  K_{\rm eff} &\equiv 
  \frac{2}{3}\left\langle\nabla^2 \varPhi + a^2 \frac{\dot{a}}{a} \dot{\varDelta}\right\rangle \\
  &= \frac{2}{3}\left\langle - a^2 \dot{H} \varDelta + a^2 H \dot{\varDelta} \right\rangle \\
  &= \frac{2}{3} a^2 H^2 \left( \frac{\mathcal{D}_+}{H}\right)^{\cdot} \langle Q_+(x_i)\rangle \\
  &= \frac{2}{3} \langle Q_+(x_i)\rangle
  \end{align}
is a constant which can be regarded as the effective curvature constant on the local domain in the averaged sense.  

Although \eqref{eq:perF2} looks similar to that obtained in \cite{kasai2019possible}, we emphasize the following advantages of our analysis in this paper over that in \cite{kasai2019possible}.
\begin{itemize}
  \item[(1)] \cite{kasai2019possible}'s result is heavily dependent on the solution of $\delta \propto a$ in the Einstein-de Sitter background. In particular all of the averaged quantities are defined and calculated directly using the growing mode solution in the Einstein-de Sitter background described in eqs.(14-17) in \cite{kasai2019possible}. So it is unclear whether it holds in any other background. In this paper, we explicitly showed that this averaged picture holds backgrounds other than the Einstein-de Sitter background, especially even if $\Lambda \neq 0$ background. \\
  \item[(2)] If we don't ignore the decaying mode of $\delta$, \cite{kasai2019possible} does not work. However, our discussion has no problem even if we consider the decaying mode. \\ 
  \item[(3)] It was unclear that \cite{kasai2019possible}'s result is valid gauges other than comoving synchronous gauge. Therefore, we explicitly showed that we can describe using the spatial average of gauge-invariant variables all the averaged density, expansion rate, and (effective) curvature constant in an inhomogeneous universe.
\end{itemize}

\section{The cosmological parameters in the nearby regions expressed by the gauge-invariant variables}

We define the global Hubble parameter as 
\begin{equation}
  H_0 \equiv \frac{\dot{a}}{a}\Bigr|_{t_0}
\end{equation}
and the global density parameters as 
\begin{equation}
  \Omega_{m} \equiv \frac{8\pi G \rho_b(t_0)}{3 H_0^2} 
\end{equation}
and 
\begin{equation}
  \Omega_{\Lambda} \equiv \frac{\Lambda}{3 H_0^2}\, , 
\end{equation}
where $\Omega_{m} + \Omega_{\Lambda} =1$ since we have assumed the flat background. These global parameters are supposed to be determined by the very large-scale and distant observations such as the cosmic microwave background. 

On the other hand, 
the cosmological parameters which are obtained from the observations in the local nearby regions are certainly determined by the local average density $\langle \varrho \rangle$, rather than by the background density $\rho_b$. 
We define the local Hubble parameter as 
\begin{equation}
  \tilde{H}_0 \equiv \frac{\dot{a}_D}{a_D}\Bigr|_{t_0} 
  = H_0 \left(1 - \frac{1}{3} f(t_0)\langle\varDelta\rangle_{t_0} \right)\, ,
\end{equation} 
where 
\begin{equation}
  f(t) \equiv \frac{d \ln \mathcal{D}_+}{d \ln a}
\end{equation}
is the growth function of the gauge-invariant density perturbation $\varDelta$, 
and the local density parameters as 
\begin{equation}
  \tilde{\Omega}_m \equiv \frac{8\pi G \langle \varrho \rangle}{3 \tilde{H}^2_0} 
  = \Omega_m \left\{1 + \left(1+\frac{2}{3} f(t_0)\right) \langle\varDelta\rangle_{t_0} \right\} 
\end{equation}
and
\begin{equation}
  \tilde{\Omega}_{\Lambda} \equiv \frac{\Lambda}{3 \tilde{H}_0^2}
  = \Omega_{\Lambda} 
  \left(1 + \frac{2}{3} f(t_0) \langle\varDelta\rangle_{t_0} \right)  \, , 
\end{equation}
which are valid up to the linear order in the gauge-invariant variable $\varDelta$. 

The local cosmological parameters coincide with the global ones if and only if $\langle\varDelta\rangle =0$.  
Otherwise, the local parameters may change. Let us show a simple estimation in the case $\Lambda = 0$, where $f(t) = 1$.  If the local nearby region is, say,  $30\%$ under dense, namely 
$\langle\varDelta\rangle_{t_0}= -0.3$, the local Hubble parameter $\tilde{H}_0$ can be $10\%$ larger than the global $H_0$. 

%%% Kasai

\section{Conclusion and Discussion}
Motivated by the Hubble tension, there have been many studies on the possible  resolutions.
One of them is the local variation of the cosmological parameters due to inhomogeneous matter distribution.
We have also studied the inhomogeneous universe by spatial averaging and obtained an interesting result on the relation between the local and global Hubble parameters which might explain the Hubble tension.

However, the question of the gauge invariance of the result is not fully understood.
In this paper we address this question.
We employ the gauge-invariant linear cosmological perturbation theory to show that the relationship between local and global cosmological parameter can be describe used by the gauge-invariant physical quantities that averaged in the local region.

It is of some interest to develop this treatment to the second order since the density contrast report by the observation of the K-band luminosity density is of the order $-0.5$.
Although we gave an argument based on the order of magnitude discussion of the cosmological Poisson equation, it is clearly not sufficient.
Another direction of this study is to consider  the possible interpretation by the inhomogeneity of the observation of m-z relation of Type Ia supernovae and CMB Power spectrum.
We hope to study this possibility in future.

\section*{Acknowledgment}

This work is supported by a Grant-in-Aid for Scientific Research from JSPS(No. 20K03937) for TF.

\vspace{0.2cm}
\noindent

%\bibliographystyle{unsrt}
%\bibliography{gauge-invariant}

\end{document}